\documentstyle[multicol,aps,epsf]{revtex} 
\tightenlines
\begin{document}
\title{ Electric field effects in STM imaging}
\author{Kurt Stokbro, Ulrich Quaade and Francois Grey}
\address{Mikroelektronik Centret, Danmarks Tekniske Universitet, 
Bygning 345\o , DK-2800 Lyngby, Denmark.}

\date{\today}
\maketitle

\begin{abstract}
 We present a high voltage extension of the Tersoff-Hamann theory of 
STM images, which  includes the effect of the
electric field between the tip and the sample. The theoretical model is based on 
first principles electronic
structure calculations and has no adjustable parameters. We use  the
method to calculate
 theoretical STM images  of  the
monohydrate Si(100)-H(2$\times$1) surface with missing hydrogen defects at $- 2$~V  and  find
an enhanced corrugation due to the electric field, in good agreement 
  with experimental images.
\end{abstract}
\begin{multicols}{2}
\narrowtext

\section{INTRODUCTION}

First principles  electronic structure calculations have become an
important tool in interpreting  STM experiments. Calculations of 
theoretical  STM images 
 are  often based on the theory by Tersoff and Hamann\cite{TeHa85},
which states that the the STM current is proportional to the local
 density of states(LDOS) of the sample. In this theory it is assumed that the
potential is flat between the tip and the sample, and the vacuum level
given by the sample work function. However, for relatively high
biases ($>2$~V), which are often used in STM experiments on semiconductor
surfaces,  the electric field strength in the tunnel region can be relatively high
and must be included in the theoretical model\cite{NeFiBr97}. 

In this paper we extend the Tersoff-Hamann 
formalism to include the electric field in the tunnel region, and apply
the theory to calculate  the corrugation of a single missing hydrogen
defect on the monohydride Si(100)-H(2$\times$1) surface. We find that the
corrugation is strongly increased by the electric field, mainly due to
polarization effects and partly due to  changes
in the tunnel barrier.

 The organization of the paper is the following:  In section II
we present the basic theory for calculating field
dependent STM images, and in section~III we show how the electric field
effect can be included in the first principles calculation. 
In section~IV we apply the formalism to calculate the corrugation of a missing
hydrogen defect on the monohydrate Si(100)-H(2$\times$1) surface and in section~V we conclude.

\section{Theory}

In this section we  present the basic theory  for calculating
field dependent STM images. The derivation  will 
follow Chen\cite{Ch93} closely.  Figure~1  shows the tunnel junction
between the tip and sample. Using the modified Bardeen approach\cite{Ch93},
 the tunnel current is given by 
\begin{eqnarray}
\label{eq:bardeen}
I & = & \frac{2 \pi e}{\hbar}\sum_{\mu \nu}
[f(\epsilon_{\nu}-e V_{\rm b})-f(\epsilon_{\mu})]|M_{\mu \nu}|^2 
\delta(\epsilon_{\mu}-\epsilon_{\nu}), \\
M_{\mu \nu}  &= & \frac{\hbar^2}{2 m} \int_{\Sigma} d\vec{S} \cdot 
(\chi^*_{\nu}\vec{\nabla}\psi_{\mu} - \psi_{\mu}\vec{\nabla}\chi^*_{\nu}
),
\label{eq:mat}
\end{eqnarray}
where the integral in Eq.~(2) is over any separating surface $\Sigma$ lying entirely
within the vacuum  region separating the two sides. The sample bias,
$V_{\rm b}$ defines the difference between tip and sample Fermi
levels, and $f(\epsilon )$ is the Fermi function. The 
modified sample(tip) wave functions $\psi_{\mu}$ ($\chi_{\nu}$) are  
 solutions to the Schr\"{o}dinger equation with   modified
sample(tip)  potential $U_s$
($U_t$). These  potentials  are given by the tunnel
potential $U$
upto the separating surface, $\Sigma$, and are equal to the vacuum
level  beyond the separating surface, thus $U=U_s+U_t$ and $U_s U_t = 0$. 
The gradient of $U$ in the tunnel region determines
the tip induced electric field, ${\bf E}={\bf \nabla} U/e$.

\begin{figure}
\begin{center}
\leavevmode
\epsfxsize=84mm
\epsffile{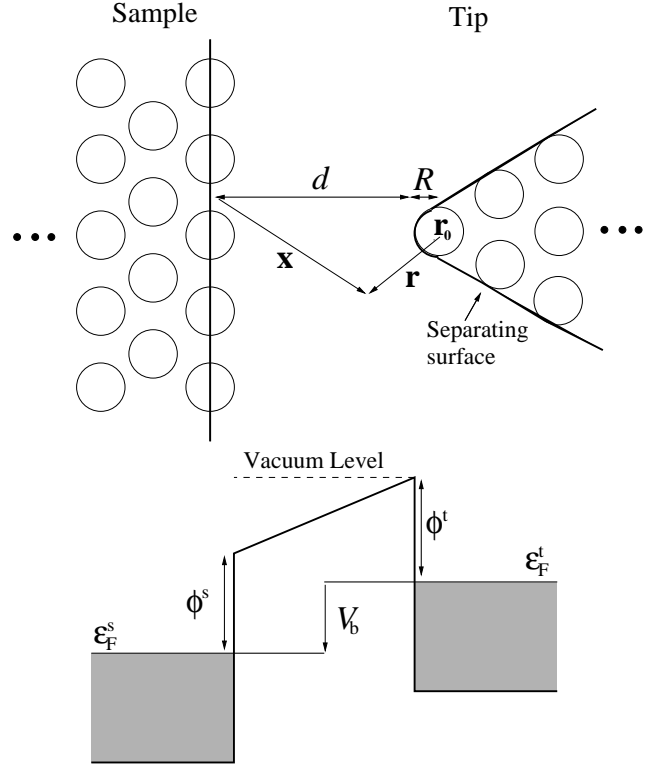}
\end{center}
\caption{ The STM tunnel region and tunnel potential. }
\end{figure}

 We assume that the tip is grounded and the
vacuum level   determined by  the tip workfunction.
Furthermore, we assume that the tunnel current is due to a single atom
at the tip apex. We
place the separating surface, $\Sigma$,  just outside the atomic
radius, $R$,  of this atom (see
Fig.~1). The position of the tip we denote by ${\bf r}_0$, and the tip
sample distance is given by $d=r_0-R$.  Since the tip potential equals
the vacuum level beyond the separating surface, it is straightforward 
 to  expand the  modified
tip wave functions in real spherical harmonics, $Y_l^m$, and obtain
\begin{eqnarray}
\chi_{\nu}({\bf r}) &=& \sum_{lm} C_{lm}^{\nu} 
k_l(\kappa_\epsilon r)/k_l(\kappa_\epsilon R) Y_l^m(\hat{r}), \\
C_{lm}^{\nu} & = & \int_{4\pi} \chi_{\nu}(R \hat{r} )
Y_l^m(\hat{r}) d\hat{r}, \\
\kappa_\epsilon &=&\sqrt{2m(\phi^t+eV_{\rm
    b}+\epsilon_F^s-\epsilon)}/\hbar,
\end{eqnarray}
where ${\bf r}$ is the  distance from the tip atom,
 $ k_l$ the spherical modified
Bessel functions, $\kappa_\epsilon$ the
inverse decay length of the electron states in vacuum, and
$\epsilon_F^s$ the Fermi level of the sample.
Following
Ref~\cite{Ch93} we  observe that   the tip Green`s function,
 defined  by $(-\hbar^2\nabla^2/2m + U_t) G({\bf r}) = 4 \pi \delta({\bf
r})$, is related to  the tip wave-function by
$k_l(\kappa_\epsilon r)Y_l^m(\hat{r})=\kappa_\epsilon^{-l-1} a_l \hat{A}_{lm} G({\bf r})$,
where the differential operators $\hat{A}_{lm}$ are  defined in
Table~1, and the coefficients $a_l$ defined by $a_{\{s,p,d\}}=\{\sqrt{1/4\pi},
  \sqrt{3/4\pi}, \sqrt{15/16\pi} \}$.
We now obtain the current
\begin{eqnarray}
\nonumber
I &=& 8 \pi^3\frac{\hbar^3 e}{m^2}\int_{\epsilon_F^s}^{\epsilon_F^s+eV_{\rm b}} \sum_{lm,\mu}\left|
\frac{ a_l \hat{A}_{lm}\psi_{\mu}({\bf
r_0})}{\kappa_\epsilon^{l+1}k_l(\kappa_\epsilon
R)}\right|^2\delta(\epsilon-\epsilon_\mu)\\
& & \times  D_{lm}(\epsilon-eV_{\rm b}) d\epsilon,
\end{eqnarray}
where we have neglected coherence between partial tip states and 
$D_{lm}(\epsilon)=\sum_\nu | C_{lm}^{\nu} |^2 \delta
(\epsilon-\epsilon_\nu )$ are  partial tip density of states per
unit volume. 

We have calculated $D_{lm}$  for a single W atom 
on a W(110) surface using $R=3$~bohr. We find that it is nearly independent of $m$,
and average  $l$ dependent values are $D_{\{s,p,d\}}(\epsilon)\approx
\{0.002
\Theta_{[-10,20]}(\epsilon)$, $0.002\Theta_{[0,20]}(\epsilon)$, $0.002\Theta_{[-4.5,3.5]}(\epsilon)\}~$eV$^{-1}$bohr$^{-3}$,  where the step function, $\Theta_{[a,b]}(x)$, is
one for $x$ in   the interval   $a<x<b$ and zero otherwise.
 Using these values  we obtain
the current 
\begin{equation}
I  =   \int_{\epsilon_F^s}^{\epsilon_F^s+eV_{\rm b}} e^{2 \kappa_\epsilon R} \sum_{lm}
B_{lm}(\epsilon-eV_{\rm b}) \rho_{lm}(d+R,\epsilon)  d\epsilon ,
\end{equation}
where distances are in bohr, energies in eV and current in Amperes.
Parameters $B_{lm}$ are defined in Table~1. 
The main quantity is the sample LDOS,
$\rho_{lm}({\bf x},\epsilon) = \sum_\mu |\hat{A}_{lm} \psi_\mu({\bf
  x})|^2 \delta(\epsilon-\epsilon_\mu)$. The wave functions
$\psi_{\mu}$ are calculated in the  external electric field from the 
tip, and  we approximate this  field by a planar electric field 
of strength $E$.
 For a
given tip-sample distance  ${ d}$, the  field strength is determined 
from the equation
\begin{equation}
U_s(d,E) = \phi^t + e V_{\rm b}+\epsilon_F^s,
\end{equation}
where $U_s$ is
the  effective sample potential in planar field $E$.  This equation assumes
 that the tip behaves as
a metallic sphere of radius $R$,  consistent with the spherical
potential-well model of the tip used in Eq.~(3)\cite{TeHa85}.

The main result of this paper, Eq.~(7),  is a high voltage generalization of the Tersoff-Hamann
expression\cite{TeHa85} for the STM current. The main differences
between Eq.~(7) and the expression by Tersoff and Hamann are the
integration over the electronic states  and the calculation of the
sample wave functions in an external electric field. We  also
include higher angular tip states\cite{Ch93}, whereas the
Tersoff-Hamann formulation is for an $s$-type state only. For the
systems we have investigated we find that $m>0$ terms are more than
one order of
magnitude smaller than $m=0$ terms and can therefore be neglected. Of
the $m=0$ states, we find  that the $l=0$ state  gives a contribution which
is twice that of  $l>0$ states. In the 
following we will only consider the $l=0$ contribution, since we have
found that this contribution best describes the experimental images we
consider. However, we  note that
occasionally we see a change in the image contrast, which might be due to
dominance of $l>0$ states for special tip geometries.

\begin{table}
\begin{tabular}{cr|cc}
$l$ &$m$ & \multicolumn{1}{c}{ $B_{lm}(\epsilon)$} & 
\multicolumn{1}{c}{ $\hat{A}_{lm}$} \\
\tableline
0&0   & $ 0.007 R^2\Theta_{[-10,20]}(\epsilon)$ & 1 \\
1&-1  & $0.02 R^4(1+\kappa_\epsilon R)^{-2} \Theta_{[0,20]}(\epsilon)$ &
$\frac{\partial}{\partial x}$  \\
1&0   &  $0.02 R^4(1+\kappa_\epsilon R)^{-2} \Theta_{[0,20]}(\epsilon)$ &
$\frac{\partial}{\partial z}$  \\
1&1   &  $0.02 R^4(1+\kappa_\epsilon R)^{-2} \Theta_{[0,20]}(\epsilon)$  &
$\frac{\partial}{\partial y}$  \\
2&-2  &  $ 0.03 R^6(3+3 \kappa_\epsilon R+\kappa_\epsilon^2 R^2)^{-2}
\Theta_{[-4.5,3.5]}(\epsilon)$ &
$\frac{\partial^2}{\partial x\partial y}$ \\
2&-1  &  $ 0.03 R^6(3+3 \kappa_\epsilon R+\kappa_\epsilon^2 R^2)^{-2}
\Theta_{[-4.5,3.5]}(\epsilon)$ &
$\frac{\partial^2}{\partial y\partial z}$ \\
2&0  &  $ 0.03 R^6(3+3 \kappa_\epsilon R+\kappa_\epsilon^2 R^2)^{-2}
\Theta_{[-4.5,3.5]}(\epsilon)$ &
$\frac{\sqrt{3}\partial^2}{\partial z^2}-\frac{\kappa_\epsilon^2}{\sqrt{3}}$ \\
2&1  &  $ 0.03 R^6(3+3 \kappa_\epsilon R+\kappa_\epsilon^2 R^2)^{-2}
\Theta_{[-4.5,3.5]}(\epsilon)$ &
$\frac{\partial^2}{\partial x\partial z}$ \\
2&2  &  $ 0.03 R^6(3+3 \kappa_\epsilon R+\kappa_\epsilon^2 R^2)^{-2}
\Theta_{[-4.5,3.5]}(\epsilon)$ &
$\frac{\partial^2 }{\partial x^2}-\frac{\partial^2  }{\partial y^2}$ \\
\end{tabular}
\caption{Definition of the parameters $B_{lm}$ and differential operators $\hat{A}_{lm}$ used
  in Eq.~(7). The step function, $\Theta_{[a,b]}(x)$, is one for $x$ in
  the interval   $a<x<b$ and zero otherwise.}
\end{table}

\section{Calculation of surface electronic structure in field $E$}

The electronic structure calculations are based on density functional
theory\cite{HoKo64,KoSh65}  using
the Generalized Gradient Approximation of Ref.~\cite{PeWa91} for
 the exchange-correlation energy. Ultra-soft pseudo potentials\cite{Va90}
including the nonlinear core correction\cite{LoFrCo82} are used
to describe hydrogen and silicon and the wave functions are
represented in a plane-wave basis set with kinetic energy cutoff 20~Ry. 
At distances larger than 4~\AA\ from the surface, the wave functions
are obtained by outward integration using the average effective
potential perpendicular to the slab\cite{Te89}.

In the following we will consider the 
 monohydrate Si(100)-H(2$\times$1)  surface, which we model by a (2$\times$1) slab
with 12 layers of silicon atoms, and a vacuum region of 10~\AA . 
We  apply an external electric field to 
 the surface, by inserting a dipole layer  in the middle of
the vacuum region\cite{NeSc92}.  The effect of mobile carriers is introduced by fixing the atoms on the  
the back surface of the slab in their bulk
 positions\cite{KrHaGrNo96}. This  gives rise to 
half-filled surface states 
 in the middle of the band gap, and this surface is therefore
 metallic.  Depending on the direction of the
 external field the surface
states accept or donate electrons. To obtain the field dependence of the  wave functions we
 calculate the  wave functions
 for two fields, $E_1$, $E_2$ , which bound the field range in the
experiment. The  wave functions at a given field, $E_1 < E <
 E_2$, are then obtained by  logarithmic interpolation between 
the wave functions at $E_1$ and $E_2$.

Figure~2 shows the effective one-electron potentials for calculations
 with external fields of $E=0.6$~V/\AA\ and $E=-0.8$~V/\AA . While the
 potential on the  back surface  is the same for both fields, the
potential on the monohydrate(front)  surface bends
 upwards for  positive fields and downwards for
negative fields, respectively.

The inset in Fig.~2 shows the band bending in the slab calculation
compared with the band bending in a n-type sample with $N_D=10^{18}$~cm$^{-3}$.
The band bending calculation is based
on standard band bending theory with non-degenerate statistics.\cite{SeGr58}
For the n-type sample the Fermi level is close to the conduction band, 
and we have the well known depletion and inversion for positive fields.
In the case of the slab calculation, mobile carriers are simulated by
the half-filled dangling-bond states on
the back surface, and since these states 
fix the Fermi level in the middle of the band gap, the band bending  is
nearly symmetric in the field. The experimental  Fermi level
is given by\cite{Sz85}
\begin{eqnarray}
\label{eq:ef}
\epsilon_F^{s} & = &\epsilon_g/2-0.49 kT +kT \log(\frac{N_D}{n_i}), \\
\epsilon_g & = &1.17 - \frac{4.73\times 10^{-4} T^2}{T+636} {\rm eV}, \\
n_i & = & 10^{16 } T^{\frac{3}{2}} e^{\epsilon_g/2kT} { \rm cm}^{-3}, 
\end{eqnarray}
where $\epsilon_g$ is the bandgap, $n_i$ the number of intrinsic
carriers and $T$ the surface temperature.

To correct for the difference in Fermi-level and band bending, $\Phi$,
between the experiment and the slab model we shift the STM voltage in
the slab model by $\Delta V_{\rm b}^+ = \tilde{\Phi}+
\tilde{\epsilon}_g-\tilde{\epsilon}_F^s- 
(\Phi+\epsilon_g-\epsilon_F^s)$ at positive bias, and 
$\Delta V_{\rm b}^- =  \tilde{\Phi}-\tilde{\epsilon}_F^s- (\Phi-\epsilon_F^s)$ at negative
bias (values with tilde are slab quantities). In this way we obtain
that  the energy window of  electronic states  which contributes to the
current  is the same in the slab model  as in the experiment.

\begin{figure}
\begin{center}
\leavevmode
\epsfxsize=84mm
\epsffile{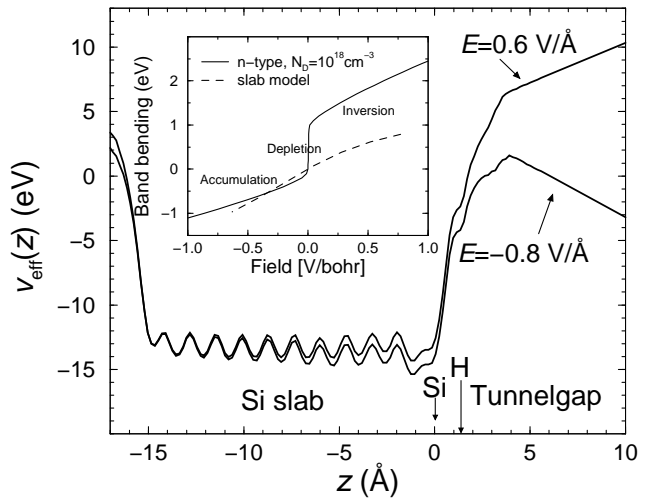}
\end{center}
\caption{\protect\small The solid lines show the average effective
potential, $v_{\rm eff}$, along the $z$
axis(perpendicular to the  slab) for external fields of 
 0.6~V/\AA\ and -0.80~V/\AA . 
 The zero of the $z$ axis is taken at the
position of the first layer Si atoms, and the Fermi level defines the
zero effective potential. The inset shows the band bending in the slab
calculation(dashed line) compared with  the band bending of a n-type sample
with $N_D=10^{18}$~cm$^{-3}$ at room temperature(solid line)\protect\cite{SeGr58}.}
\end{figure}

\section{The STM corrugation of Si(100)-H(2$\times$1)}

In Fig.~3 we show a typical STM filled state image of the monohydrate
Si(100)-H(2$\times$1) surface. The bright vertical stripes originate
 from the hydrogen passivated silicon
dimer rows, and the white spot originates from a silicon dangling bond
due  to a single missing hydrogen defect. Below the image we show the
corrugation across the 
defect(solid line), compared with the simulated STM image of an
$s$-state tip including field
effects(dashed line) and without field effects(dotted line). In the
range -5~\AA$<x<$10~\AA\  the theoretical curves were obtained using a
c(4$\times$4) cell with a single missing hydrogen defect, and outside this
range using a (2$\times$1) cell. We see that the corrugation of the
defect is well described in the field dependent calculation, while it
is less than half the value when the field is not included. The larger corrugation of the defect in
 the field dependent calculation is mainly 
due to polarization  of the dangling bond. 
 Away from
the defect the corrugation in both calculations is 
less than in the experiment.  Calculations using $p$- or $d$-state tips do
not give better agreement, and we suggest that the larger experimental
corrugation away from the defect might be  due to thermal vibrations
on the surface. 

\begin{figure}
\begin{center}
\leavevmode
\epsfxsize=84mm
\epsffile{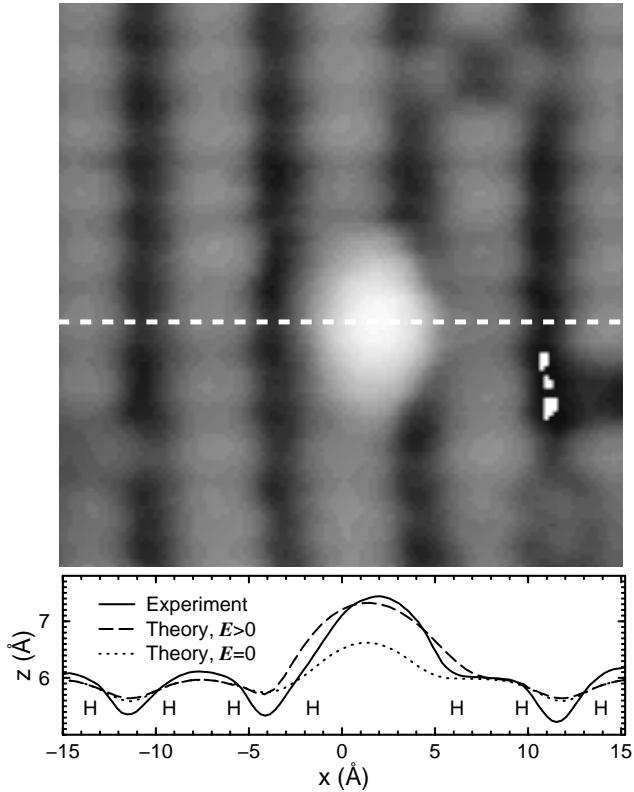}
\end{center}
\caption{Filled state STM image of a single hydrogen defect on the
  Si(100)-H(2$\times$1) surface, recorded with $V=-1.6$~V and
  $I=1$~nA. The  plot shows the corrugation across the defect(solid 
  line), together with the theoretical corrugation including electric
  field effects(dashed line) and without field effects(dotted line).}
\end{figure}

\section{Conclusions}
We have presented a high voltage extension of the Tersoff-Hamann model of STM
images, which includes the  electric field between tip and sample. We
have applied the model to describe the corrugation of a single missing
hydrogen defect, and find good agreement with experiment when field
effects are included. At low voltages $|V_{\rm b}| < 3$~V, the field induced
change of the corrugation is mainly due to polarization. 
For higher voltages $|V_{\rm b}|>3$~V, as used in many atom manipulation
 experiments, the  field  also has a pronounced effect on the tunnel
 barrier, and we hope that the present work may prove useful for the
 analysis of such experiments.

\end{multicols}
\end{document}